# Combining dissimilarity measure for the study of evolution in scientific fields

Lukun Zheng &Yuhang Jiang

*Department of Mathematics, Western Kentucky University, Bowling Green, USA.*

Lukun Zheng, email: lukun.zheng@wku.edu.

# Combining dissimilarity measure for the study of evolution in scientific fields

The evolution of scientific fields has been attracting much attention in recent years. One of the key issues in evolution of scientific field is to quantify the dissimilarity between two collections of scientific publications in literature. Many existing works study the evolution based on one or two dissimilarity measures, despite the fact that there are many different dissimilarity measures. Finding the appropriate dissimilarity measures among such a collection of choices is of fundamental importance to the study of scientific evolution. In this article, we develop a new measure of the evolution combining twelve keyword-based temporal dissimilarities of the scientific fields using the method of principal component analysis. To demonstrate the usage of this new measure, we chose four scientific fields: environmental studies, information science & library science, mechanical informatics, and religion. A database consisting of 274453 bibliographic records in these four chosen fields from 1991 to 2019 are built. The results show that all these four scientific fields share an overall decreasing trend in evolution from 1991 to 2019 and different fields exhibits different evolution patterns during different time periods.

Keywords: dissimilarity measures; measure of evolution; principal component analysis.

**Introduction**

In biology, evolution is change in the heritable characteristics of biological populations over successive generations (Arnold, 1997; Ohno, 2013). The scientific theory of evolution by natural selection began with Charles Darwin's On the Origin of Species published in 1859 (Darwin, C., & Bynum, W. F., 2009). Evolution as a scientific theory has been used in many other disciplines as well, including medicine (Grunspan et al, 2018; Nesse, 2008) , psychology (Panksepp, J., & Panksepp, J. B., 2000; Langs, 2019), anthropology (Bateson, 2000; Tang, 2020), forensics (Khan, M. Z., Mishra, A., & Khan, M. H., 2020; Poore, J., Flores, J. C., & Atkison, T., 2013), agriculture (Mueller, U. G., Rehner, S. A., & Schultz, T. R., 1998; Fordham, 2019)and other social-cultural applications (Donald, 1991; Giavazzi, F., Petkov, I., & Schiantarelli, F., 2019).

This work aims at studying the evolution of scientific fields by investigating their developmental trends shown in scientific literature. The temporal evolution of scientific research can be observed in retrospective studies in many fields. What's more, the "evolution map" of scientific fields helps us understand the nature of scientific development and the relative importance of different topics or publications (Tang, X., Yang, C., & Song, M., 2013). Innovations and scientific breakthroughs keep on emerging, leading to new or improved technology and scientific findings, which, then, shape new developmental trends in various research areas. In this article, we study how scientific fields evolve over time.

There has been a growing interest in developing topic detection and tracking methodologies to study how topics evolve. Cai et al. (2019) defined three kinds of event relationships: temporal, content dependence, and event reference, that can be used to identify to what extent a component event is dependent on another in the evolution of a target event. Gaul & Vincent (2017) introduced an information clustering approach based on content-(dis)similarity of the underlying textual material and graph-theoretical considerations to deal with the network of relationships between content-similar topics. Zhou et al. (2017) reviewed the extant research on topic evolution map based on text and cross-media corpora over the past decade. Chen et al. (2017) examined how research topics evolve by analyzing the topic trends, evolving dynamics, and semantic word shifts in the IR domain using a data set on information retrieval (IR) publications. They found that the evolution of a major topic usually follows a pattern from adjusting status to mature status, and sometimes with re-adjusting status in between the evolving process. Many other works study topic evolution using a generative modles. Xu et al. (2018) adopted extracted keywords to investigate the features of interdisciplinarity development, as well as the distinct roles that different participating domains play in various periods, and detect potential barriers among domains. Xu et al. (2019) introduced a research front detection and topic evolution approach based on graph theory utilizing the topological structure and the PageRank algorithm. Xu et al. (2019) utilzed several machine learning models detect and foresight the emerging research topics. Although these works showed great effectiveness in topic modelling within a certain information domain, their main emphases were on topics modeling instead of transitional changes in scientific fields.

Evolution in scientific fields has been drawing attention among researchers and scientists in recent years. One of the main issues in evolution of scientific field is to quantify the dis/simmilarity between two collections of scientific publications in literature. Vargas-Quesada et al. (2010) introduced a graphic reprentation of the intellectual structures in the form of scientograms using scientometric information such as cocitation network from a certain scientific domain. Their approach allows one to detect patterns and tendencies of scinetific evolution in a scientific domain through network visualizations. Tang et al. (2013) incoprated content similarity and coauthorship networkd similarity to quantify the similarities among different research domains. Dias et al. (2018) used an information-theoretic measure of linguistic similarity to investigate the organization and evolution of scientific fields based on the 20,000 most frequent words from the abstracts of the papers considered excluding a list of stop words. There are also works studying the evolution of citations in scientific fields. Finardi (2014) studied the time evolution of mean received citations on a sample of journals from two ISI subject categories. Frank et al. (2019) used the Microsoft Academic Graph to study the bibliometric evolution of AI research and its related fields from 1950 to 2019. The problem with these works is that they only adopted few (mostly just one single) dissimilarity measures in their study, ignoring the fact that there are many such measures, and hence failed to discuss the comparison and choices of such measures. There are a number of dissimilarity measures encountered in many different areas such as biology, computer science, mathematics, psychology, statistics, etc. Finding the appropriate measures among such a collection of chioces is of fundamental importance to pattern classification, clustering, and information retrieval problems, see Duda, Hart, and Stork (2012). There have been such endeavors in many different fields (Wei, 2018; Cleasby et al. 2019; McCulloch & Wagner, 2020; Jain, Mahara, & Tripathi, 2020).

In this paper, we base our study on a set of twelve dissimilarity measures and propose a new approach to quantify the speed of evolution in a scientific field using the method of principle component analysis. We conduct a thorough analysis and comparison of different textual dissimilarity measures in the study of scientific evolution, while most of the existing works only consider few (mostly just one) such textual similarity measure. Our methodology is based on the probability distributions derived from the keywords observed in a collection of publications. We introduce an unsupervised methodology to analyze the progress of scientific literature and perform a systematic anlaysis on how different scientific fields evolve and compare their evolution speeds over the last three decades.

Another feature of our work is that we focus on scientific fields, while most of the previous studies focus on research topics. Each scientific field is defined by papers of the same category as classified by Web of Science (Web of Science). In this paper, the scientific field consists of closely related research topics and can be represented by a collection of scientific literature within that field. For instance, artificial intelligence is a research field. Within this fields, there are many different research topics such as deep learning, computer vision, natural language processing, recommender systems, robotics, and so on. Similarly, environmental studies can be another scientific field, which consists of research topics such as arid lands, climate adaptation and sustainability, ecohydrology and biogeochemistry, invasive species, plant and soil ecology, watershed management, wildlife ecology, and so on.

In this study, we focus on four scientific fields namely, environmental studies (ES), information science & library science (ILS), mechanical informatics (MI), and religion (RL), due to their different evolution patterns in the past three decades. The data used in this paper was retrieved from the Web of Science, which gives access to multiple databases that reference cross-disciplinary research for in-depth exploration of specialized sub-fields. Furthermore, in this study, the field tag WC(=Web of Science Categories) was used to create the query. The data set contains all available factors including the title, author(s), keywords, abstract, citation information, etc.

The rest of the paper is organized as follows. The Methodology section introduces the research methods applied to address the proposed questions. We conducted intensive experimental studies and discuss the results in the Experimental

Study section. Finally, we conclude our work with a summary of the current work and a discussion of future work in the Conclusions section.

**Methodology**

The empirical analysis performed in the present work is based on a collection of bibliographic records of English articles from the four scientific fields mentioned above over the 1991-2019 period. We chose only English articles since English is the absolute dominant language type in scientific literature and the inclusion of article in other languages may cause potential mistakes caused by translation, which may lead to unconvincing results. Due to the facts that author keywords are included in records of articles from 1991 forward in Web of Science and the data in 2020 were not completely available at the time of data collection, the data set only covered data from 1991 to 2019. Furthermore, because publication year naturally partitions the data set and our data span 29 years (1991-2019), we divide the dataset into 29 time intervals. More specifically, we represent a scientific field by a collection $B$ of bibliographic records. This collection $B$ is then partitioned based on the publication year into 29 yearly data set $B_t$, for $t = 1, 2, \ldots, 29$ with $t = 1$ representing the year 1991, $t = 2$ representing the year 1992, and so on.

*Keyword distributions*

We extract all the keywords from the data set $B_t$ for a scientific field in the $t$-th year and then obtain the frequency distribution of all the keywords. Assume that there are, in total, $v_t$ different keywords: $w_1, w_2, \cdots, w_{v_t}$ with frequencies $f^t_{w_1}, f^t_{w_2}, \cdots, f^t_{w_{v_t}}$. Let $f^t = f^t_{w_1} + f^t_{w_2} + \cdots, + f^t_{w_{v_t}}$ be the total frequency of all the keywords extracted from $B_t$. We obtain the relative frequency table of the keywords for year $t$ and denote it as $P_t \equiv \{p^t_{w_i}\}_{i=1}^{v_t}$ with $p^t_{w_i} = f_{w_i}/f$ for $i = 1, \ldots, v_t$. This table is called the keyword distribution for year $t$ for this scientific field. Similarly, the keyword distribution for another year $s$ can also be obtained and we denote it as $P_s \equiv \{p^s_{u_i}\}_{i=1}^{v_s}$, where $u_1, u_2, \cdots, u_{v_s}$ are the distinct keywords used in year $s$ and $v_s$ is the total number of distinct keywords.

Now that we have the keyword distributions $P_t$ and $P_s$ of both years, how can we measure the dissimilarity between them? First, in order to measure the dissimilarity between these two keyword distributions $P_t$ and $P_s$, we need to base these two distributions on the same set of keywords. That is, we need to modify these distributions $P_t$ and $P_s$ so that they are different distributions on the same set of keywords, allowing zero probability values. In order to do so, it is important to note that there might be many common keywords in these two years, considering that they are the same scientific field. That is, $w_1, w_2, \cdots, w_{v_t}$ and $u_1, u_2, \cdots, u_{v_s}$ may share common keywords even though we denote them differently. Assume that there are $c_{t,s}$ common keywords among them, then the total number of distinct keywords in both years is $v_{t,s} = v_t + v_s - c_{t,s}$. Let's denote these $v_{t,s}$ keywords as $kw_1, kw_2, \cdots, kw_{v_{t,s}}$. Among these keywords, some are only from year $t$, some are only from year $s$, and the rest are shared by both years. We will update $P_t$ and $P_s$ using these $v_{t,s}$ keywords: $kw_1, kw_2, \cdots, kw_{v_{t,s}}$ as follows:

1. For each keyword $kw_i$ in $kw_1, kw_2, \cdots, kw_{v_{t,s}}$, we put $p^t_{kw_i} = p^t_{w_{i\prime}}$, if it is from $w_1, w_2, \cdots, w_{v_t}$ and the keyword $kw_i$ is the same as the keyword $w_{i\prime}$, for some $i'$ in $1, \ldots, v_t$. We put $p^t_{kw_i} = 0$ if $kw_i$ is not from the list $w_1, w_2, \cdots, w_{v_t}$. Then the distribution $P_t$ is then updated as $P_t = \{p^t_{kw_i}\}_{i=1}^{v_{t,s}}$, a distribution on the $v_{t,s}$ keywords $kw_1, kw_2, \cdots, kw_{v_{t,s}}$.

2. Similarly, for each keyword $kw_i$ in $kw_1, kw_2, \cdots, kw_{v_{t,s}}$, we put $p^s_{kw_i} = p^s_{u_{i\prime}}$, if it is from $u_1, u_2, \cdots, u_{v_s}$ and the keyword $kw_i$ is the same as the keyword $u_{i\prime}$, for some $i'$ in $1, \ldots, v_s$. We put $p^s_{kw_i} = 0$ if $kw_i$ is not from the list $u_1, u_2, \cdots, u_{v_s}$. Then the distribution $P_s$ is then updated as $P_s = \{p^s_{kw_i}\}_{i=1}^{v_{t,s}}$, a distribution on the $v_{t,s}$ keywords $kw_1, kw_2, \cdots, kw_{v_{t,s}}$.

A rationale for how to construct such keyword distributions is best explained by working though a hypothetical example. Assume that Table 1 gives the keyword distribution for year $t$ and Table 2 gives the keyword distribution for year $s$. From these tables, we see that there are $v_t = 4$ different keywords for year $t$ and $v_s = 5$ different keywords for year $s$. There are $c_{t,s} = 3$ common keywords for these two years. Hence the total number of distinct keywords is $v_{t,s} = v_t + v_s - c_{t,s} = 6$. The keywords distributions in Table 1 and 2 can be updated using the above procedure and the resulting distributions are shown in Table 3.

TABLE 1. Keywords distribution for year $t$.

| **Keywords** | Neural network | Pattern recognition | Deep learning | Similarity |
|---|---|---|---|---|
| **Frequency** ($f^s_{w_i}$) | 18 | 14 | 22 | 16 |
| **Relative frequency** ($p^s_{w_i}$) | 18/70 | 14/70 | 22/70 | 16/70 |

TABLE 2. Keywords distribution for year $s$.

| **Keywords** | Artificial Intelligence | Pattern recognition | Deep learning | Robotics | Similarity |
|---|---|---|---|---|---|
| **Frequency** ($f^s_{u_i}$) | 25 | 30 | 15 | 10 | 10 |
| **Relative frequency** ($p^t_{w_i}$) | 25/90 | 30/90 | 15/90 | 10/90 | 10/90 |

TABLE 3. Updated keywords distributions for year $t$ and year $s$.

| Keywords | Artificial Intelligence | Neural network | Pattern recognition | Deep learning | Robotics | Similarity |
|---|---|---|---|---|---|---|
| Updated distribution for year $t$ | 0 | 18/70 | 14/70 | 22/70 | 0 | 16/70 |
| Updated distribution for year $s$ | 25/90 | 0 | 30/90 | 15/90 | 10/90 | 10/90 |

*Dissimilarity Measures*

Now that these two updated keyword distributions are on the same set of $v_{t,s}$ keywords $kw_1, kw_2, \cdots, kw_{v_{t,s}}$, we can consider the dissimilarity measures between them. Dissimilarity measures should satisfy the following conditions: 1) symmetric $d(x,y) = d(y,x)$, 2) non-negative $d(x,y) \geq 0$ and 3) $d(x,x) = 1$, according to Webb, (2003). There have been many efforts made in finding the appropriate dissimilarity measures in many areas. While most of the previous works uses individual dissimilarity measures, we propose to use a mixture of 12 dis/similarity measures between keyword distributions to quantify the evolution between two years. These 12 dissimilarity measures shown in Table 4 are commonly used measures belonging to different families according to Cha, (2007).

TABLE 4. Dissimilarity measures between two probability distributions $P = \{p_i\}_{i=1}^d$ and $Q = \{q_i\}_{i=1}^d$ with $0 \leq p_i, q_i \leq 1$ and $\sum_{i=1}^d p_i = \sum_{i=1}^d q_i = 1$.

| Dis/Similarity Measure | Expression | |
|---|---|---|
| **Canberra** | $d_1 = d_{Can} = \sum_{i=1}^d \frac{|p_i - q_i|}{p_i + q_i}$ | (1) |
| **Clark** | $d_2 = d_{Clk} = \sqrt{\sum_{i=1}^d \left(\frac{p_i - q_i}{p_i + q_i}\right)^2}$ | (2) |
| **Cosine** | $d_3 = d_{Cos} = 1 - \frac{\sum_{i=1}^d p_i q_i}{\sqrt{\sum_{i=1}^d p_i^2}\sqrt{\sum_{i=1}^d q_i^2}}$ | (3) |
| **Czekanowski** | $d_4 = d_{Cze} = \frac{\sum_{i=1}^d |p_i - q_i|}{\sum_{i=1}^d (p_i + q_i)}$ | (4) |
| **Euclidean $L_2$** | $d_5 = d_{Euc} = \sqrt{\sum_{i=1}^d |p_i - q_i|^2}$ | (5) |
| **Jesen-Shannon** | $d_6 = d_{JS} = \frac{1}{2}\sum_{i=1}^d \left[p_i \ln\left(\frac{2p_i}{p_i + q_i}\right) + q_i \ln\left(\frac{2q_i}{p_i + q_i}\right)\right]$ | (6) |
| **Kulczynski** | $d_7 = d_{Kul} = \frac{\sum_{i=1}^{d}|p_i - q_i|}{\sum_{i=1}^{d} \max(p_i, q_i)}$ | (7) |
| **Lorentzian** | $d_8 = d_{Lor} = \sum_{i=1}^{d} \ln(1 + |p_i - q_i|)$ | (8) |
| **Manhattan $L_1$** | $d_9 = d_{Man} = \sum_{i=1}^{d}|p_i - q_i|$ | (9) |
| **Probabilistic Symmetric $\chi^2$** | $d_{10} = d_{PS} = 2 \sum_{i=1}^{d} \frac{(p_i - q_i)^2}{p_i + q_i}$ | (10) |
| **Soergel** | $d_{11} = d_{Soe} = \frac{\sum_{i=1}^{d}|P_i - Q_i|}{\sum_{i=1}^{d} \max(P_i, Q_i)}$ | (11) |
| **Squared-Chord** | $d_{12} = d_{SC} = \sum_{i=1}^{d}(\sqrt{P_i} - \sqrt{Q_i})^2$ | (12) |

*Principal Component Analysis*

From Table 4, we see that these chosen dissimilarity measures have complex, direct or indirect relations among them. Which measures should we use for scientific evolution? Instead of using individual such measures like what many existing works did, we propose to use principal component analysis to convert this set of dissimilarity measures into mutually independent principal components. Principal component analysis (PCA) is a common statistical tool in modern data analysis - in diverse fields from economics to computer science - because it is a simple, non-parametric method to extract relevant information from complex data sets. PCA reduces a complex data set to a lower dimension to reveal the sometimes hidden, simplified structures that often underlie it, see Shlens, (2014) and Zheng & Zheng (2020).

Over the period 1991-2019, we construct the keyword distributions for each year. The dissimilarity measures in Table 4 are calculated for each successive pair of years using the following procedure:
1. Obtain keyword distributions for year $t$ for $t = 1, ..., 29$ with $t = 1$ representing the year 1991, $t = 2$ for the year 1992, and so on.
2. Obtain the updated keyword distributions for each successive pairs of years, so that these distributions are based on the same set of keywords as instructed in above.
3. Based on the updated keyword distributions, calculate the values of the dissimilarity measures shown in Table 4.

Using the procedure above, we obtain a 28-by-12 dissimilarity dataset for each of the four chosen scientific fields since we have 29-1=28 different successive pairs of years and 12 different dissimilarity measures under consideration. These four datasets will be combined to a 112-by-12 dissimilarity data set containing the dissimilarity measurements for all the four scientific fields. PCA will be applied to this combined dataset to reduce this highly correlated complex data set into 12 mutually independent principal components. We will use the first component which accounts for the most amount of the variance in the dissimilarity data set. The first principal component (PC1) is a linear combination of these 12 dissimilarity measures. More specifically, let $P_t = \{p_{kw_i}^t\}_{i=1}^{v_{t,s}}$ and $P_{t+1} = \{p_{kw_i}^{t+1}\}_{i=1}^{v_{t,s}}$ be the updated keyword distributions for year $s$ and year $t$, respectively. Then the first principal component is given by

$$PC1(P_t, P_{t+1}) = \sum_{i=1}^{12} a_i d_i(P_t, P_{t+1}) \quad (13)$$

where $a_i$ are the coefficients obtained from principal component analysis and $d_i(P_s, P_t)$ are the dissimilarity measures between these two keyword distributions shown in Table 4. We will then transform these values of the first principal component using a translation by subtracting their minimum value so that the obtained values are all non-negative and can be used as the dissimilarity measurements. That is, the dissimilarity between two successive pair of years is defined to be

$$d(P_t, P_{t+1}) = PC1(P_t, P_{t+1}) - \min_t PC1(P_t, P_{t+1}) \quad (14)$$

*The Speed of Evolution*

The speed of evolution can be measured as the amount of dissimilarity per year over a period of several years. We discussed how to measure the dissimilarity between any two keyword distributions. Now we will consider the temporal evolution of a scientific field represented by the collection of scientific literature within that field. We will use the following algorithm to quantify the amount of evolution over the years from year $t_1$ to year $t_2$ with $1 \leq t_1 < t_2 \leq 29$:

1. Obtain keyword distributions for year $t$ for $t = t_1, t_1 + 1, \ldots, t_2$.
2. Obtain the updated keyword distributions $P_t \text{ and } P_{t+1}$ for each successive pairs of years, so that these distributions are based on the same set of keywords as instructed in above.
3. Calculate the dissimilarity measure $d(P_t, P_{t+1})$ shown in Equation (14).
4. The amount of evolution is the summation of all the dissimilarity values $d(P_t, P_{t+1})$ within the time period from year $t_1$ to year $t_2$. That is,

$$D(t_1, t_2) = \sum_{t=t_1}^{t_2-1} d(P_t, P_{t+1}) \quad (15)$$

Then average speed of evolution over the time period from year $t_1$ to year $t_2$ is given as

$$V(t_1, t_2) = \frac{D(t_1, t_2)}{t_2 - t_1} = \frac{\sum_{t=t_1}^{t_2-1} d(P_t, P_{t+1})}{t_2 - t_1} \quad (16)$$

**Experimental Study**

As described in previous section, the purpose of this article is to evaluate the evolution of scientific fields based on dissimilarity measures. The dissimilarity of two keyword frequency distributions indicates how far the research direction has been changed. By analyzing the publications from 4 different scientific fields in the past 28 years, it allows us to demonstrate different patterns of evolution among these four scientific fields.

*Descriptive Statistics of the Data*

Our data set consists of 274453 bibliographic records, of which 129,248 are from the scientific field of environmental study (ES), 58,358 are from the scientific field of medical informatics (MI), 9,216 are from the scientific field of religion (RE), and 85,072 are from the scientific field of information science and library science (ILS). It is worthy to mention that some bibliographic records may belong to multiple scientific fields. For instance, one paper may be listed as one from both medical informatics (MI) and information science and library science (ILS).

Table 5 presents some relevant statistics about the data set for these four scientific fields during different time periods (1991-2000, 2001-2010, 2011-2019, and all years). The number of articles varies from field to field: with the largest number (129248) of articles in Environmental Science and the smallest number (9216) in Religion. The number of keywords represents the total number of keywords used in all the articles in a certain field within a given time period, while the number of distinct keywords represents the number of different keywords used in all the articles in a certain filed within a given time period. Although these two numbers are closely related to the number of articles, a larger number of articles doesn't necessarily lead to a larger number of keywords (or distinct keywords). For instance, the number of articles in Information Science and Library Science is 26714 larger than that in Medical Informatics, while the number of keywords and the number of distinct keywords are both smaller than those in Medical Informatics, respectively. The table also gives the average number of keywords per article in each field during different time periods. We see that Medical Informatics has the largest average number of keywords per article during all these four different time periods, while Information Science and Library Science has the smallest average during all these four different time periods.

Figure 1 presents the yearly statistics of the data for these four chosen scientific fields: Environmental Science (ES), Information Science and Library science (ILS), Medical Informatics (MI), and Religion (RE). Figure 1 (a) presents the yearly number of articles published in each of the four chosen fields, from which we can see that ES experienced the rapidest growth over these years while RE experienced the slowest growth. The other two fields are very similar in terms of growing pattern of the number of articles from 1991-2019. The yearly number of keywords and distinct keywords show similar growing patterns for each scientific field, see Figure 1 (b) and (c). Figure 1 (d) shows the yearly number of new keywords (in thousands). Note that new keywords in each year only includes those keywords which are first seen in that year and hence, never appear in previous years during the time period 1991-2019. We do not have the number of new keywords for 1991 and the figure shows the yearly values starting from 1992 since there is no records of keywords before 1991 in our data.

TABLE 5. Relevant statistics of the data set for these four scientific fields during different time periods (1991-2000, 2001-2010, 2011-2019, and all years).

|  | 1991-2000 | 2001-2010 | 2010-2019 | All Years |
|---|---|---|---|---|
| *Environmental Science* | | | | |
| # of Articles | 15072 | 30010 | 84166 | 129248 |
| # of Keywords | 22013 | 106565 | 456561 | 585139 |
| # of Distinct Keywords | 15835 | 62606 | 213236 | 291677 |
| # of Keywords/Article | 1.46 | 3.55 | 5.42 | 4.53 |
| *Information Science and Library Science* | | | | |
| # of Articles | 23548 | 26669 | 34855 | 85072 |
| # of Keywords | 13469 | 58526 | 161217 | 233212 |
| # of Distinct Keywords | 9649 | 33669 | 82416 | 125734 |
| # of Keywords/Article | 0.57 | 2.19 | 4.63 | 2.74 |
| *Medical Informatics* | | | | |
| # of Articles | 11530 | 15867 | 30961 | 58358 |
| # of Keywords | 26333 | 69290 | 169774 | 265397 |
| # of Distinct Keywords | 18033 | 43164 | 85911 | 147108 |
| # of Keywords/Article | 2.28 | 4.37 | 5.48 | 4.55 |
| *Religion* | | | | |
| # of Articles | 1873 | 2435 | 4908 | 9216 |
| # of Keywords | 3185 | 9718 | 24703 | 37606 |
| # of Distinct Keywords | 2406 | 7041 | 15765 | 25212 |
| # of Keywords/Article | 1.70 | 3.99 | 5.03 | 4.08 |

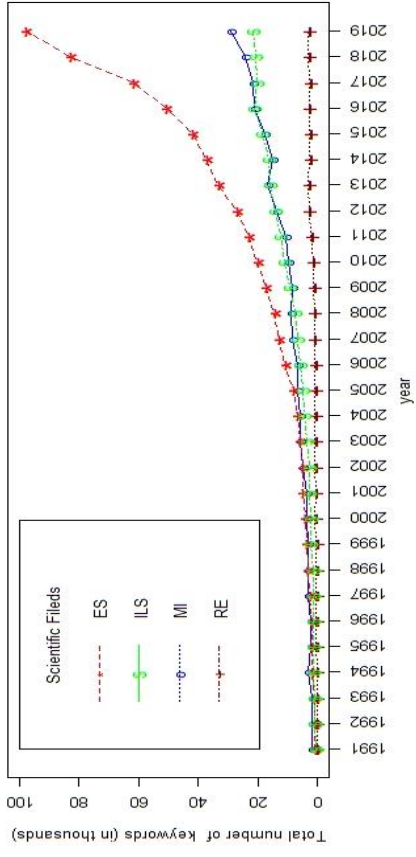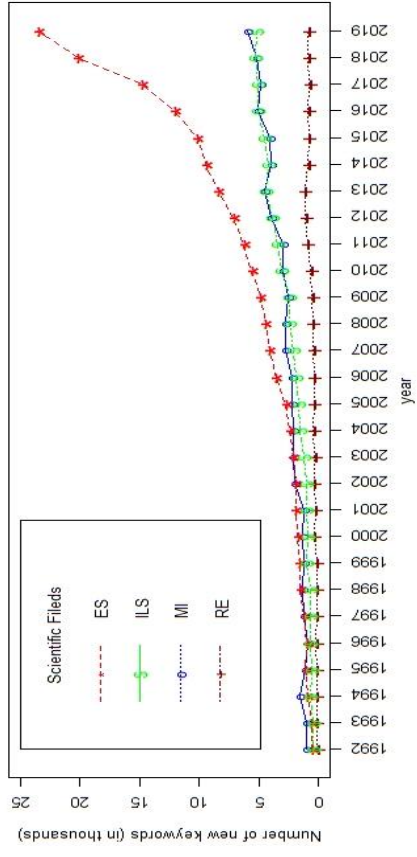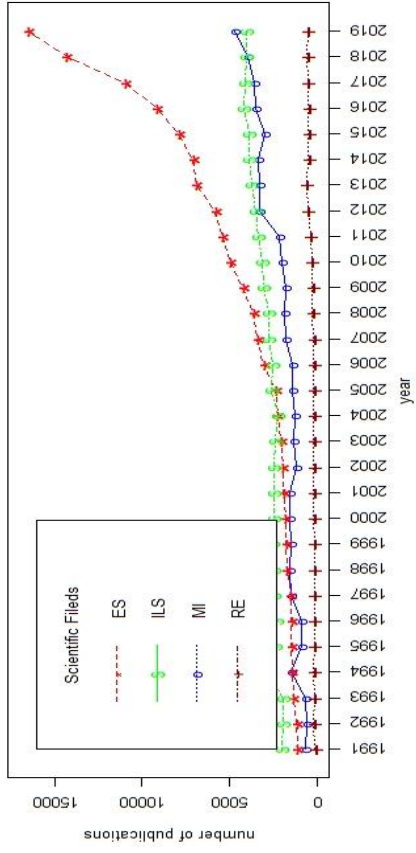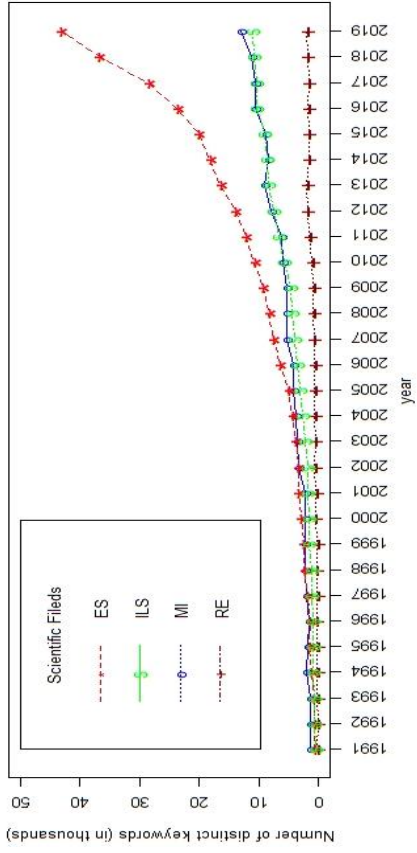

Figure 1. Yearly statistics of the data for these four chosen scientific fields: Environmental Science (ES), Information Science and Library science (ILS), Medical Informatics (MI), and Religion (RE). (a). Yearly number of articles; (b). Yearly number of keywords (in thousands); (c). Yearly number of distinct keywords (in thousands); (d). Yearly number of new keywords (in thousands).

*The Comparison of Dissimilarity Measures*

As we discussed above, there are many different dissimilarity measures and the choices of dissimilarity measures to use is of great importance in many topics. Here we compare the dissimilarity measures used in this paper based on their values between the keyword distributions of each successive pair of years from 1991 to 2019.

For each scientific field, we obtain the collection of articles and categorized them by their years of publication, from which we collected all the keywords for each year and obtained the keyword distribution $P_t$ for each year $t$ from 1991 to 2019. Afterwards, we obtain the values of the 12 dissimilarity measures shown in Table 4 based on the two keyword distributions $P_t$ and $P_{t+1}$ for each successive pair $(t, t+1)$ of years, where $t = 1991, \ldots, 2018$.

Table 6 shows the summary statistics about the dissimilarity measurements between the keyword distributions for each of these 28 successive pairs of years from 1991-2019 in the field of Environmental Science. We can see that these dissimilarity measures have different scales, with the Canberra measurements ranging from 1092.777 to 60017.153 while the Euclidean measurements ranging from 0.005 to 0.058.

Figure 2 shows the scatterplot matrix among the Clark, Czekanowski, Jensen-Shannon, Lorentzian, and Probabilistic Symmetric $\chi^2$ measurements between the keyword distributions for each of these 28 successive pairs of years from 1991-2019 in the field of Environmental Science. They have a strong linear correlation with each other with all the correlation coefficients having an absolute value over 0.9. We also see that the Clark measurements are negatively correlated with the other four measurements, while the other four measurements are all positively correlated with each other.

From Table 6 and Figure 2, we see that these dissimilarity measures have different scales and some of them may even negatively correlated. If we base our study on only one or few dissimilarity measures and ignore the fact that there are many such measures, we may reach disputable, even sometimes contradictory, conclusions on the same problem. Finding the appropriate measures among such a collection of choices is of fundamental importance to many areas such as pattern classification, clustering, and information retrieval problems. In this work, we present a method based on a class of different dissimilarity measures numerically using PCA to find a combination of these dissimilarity measures as a measure of dissimilarity.

TABLE 6. Summary statistics about the dissimilarity measurements between the keyword distributions for each of these 28 successive pairs of years from 1991-2019 in the field of Environmental Science.

| Methods | Min. | 1st Qu. | Median | Mean | 3rd Qu. | Max. |
|---|---|---|---|---|---|---|
| **Canberra** | 1092.777 | 3822.760 | 8270.484 | 14633.649 | 20533.538 | 60017.153 |
| **Clark** | 32.855 | 61.253 | 89.745 | 105.122 | 141.306 | 241.643 |
| **Cosine** | 0.038 | 0.095 | 0.286 | 0.293 | 0.446 | 0.709 |
| **Czekanowski** | 0.494 | 0.561 | 0.661 | 0.652 | 0.738 | 0.825 |
| **Euclidean** | 0.005 | 0.010 | 0.017 | 0.021 | 0.027 | 0.058 |
| **Jensen-Shannon** | 0.282 | 0.330 | 0.405 | 0.400 | 0.467 | 0.539 |
| **Kulczynski_d** | 0.914 | 1.766 | 3.167 | 3.591 | 4.883 | 8.866 |
| **Lorentzian** | 0.988 | 1.121 | 1.322 | 1.303 | 1.475 | 1.648 |
| **Manhattan** | 0.988 | 1.121 | 1.322 | 1.303 | 1.476 | 1.649 |
| **Probabilistic Symmetric $\chi^2$** | 1.666 | 1.943 | 2.375 | 2.348 | 2.729 | 3.144 |
| **Soergel** | 0.661 | 0.719 | 0.796 | 0.784 | 0.849 | 0.904 |
| **Squared-Chord** | 0.797 | 0.933 | 1.150 | 1.137 | 1.332 | 1.538 |

*Principle Component Analysis and Evolution Measurements*

With twelve strongly correlated dissimilarity measures, we use principle component analysis to find a linear combination of these dissimilarity measures which retains the most variance present in the original measurements. Figure 3 presents the scree plot of the principle component analysis for these 12 dissimilarity measures, which displays the proportion of variances retained by all principle component in a line plot from the largest to the smallest. We see that the first principle component accounts for over 80% of the total variance present in the dissimilarity measurements. We use the value of the first principle component as a measure of evolution between two successive years.

Figure 4 presents the evolution measurements for each successive pair of years from 1991 to 2019 in these four chosen scientific fields. There are several observations from this figure. First, we see that there is a decreasing trend for all fields over these years.

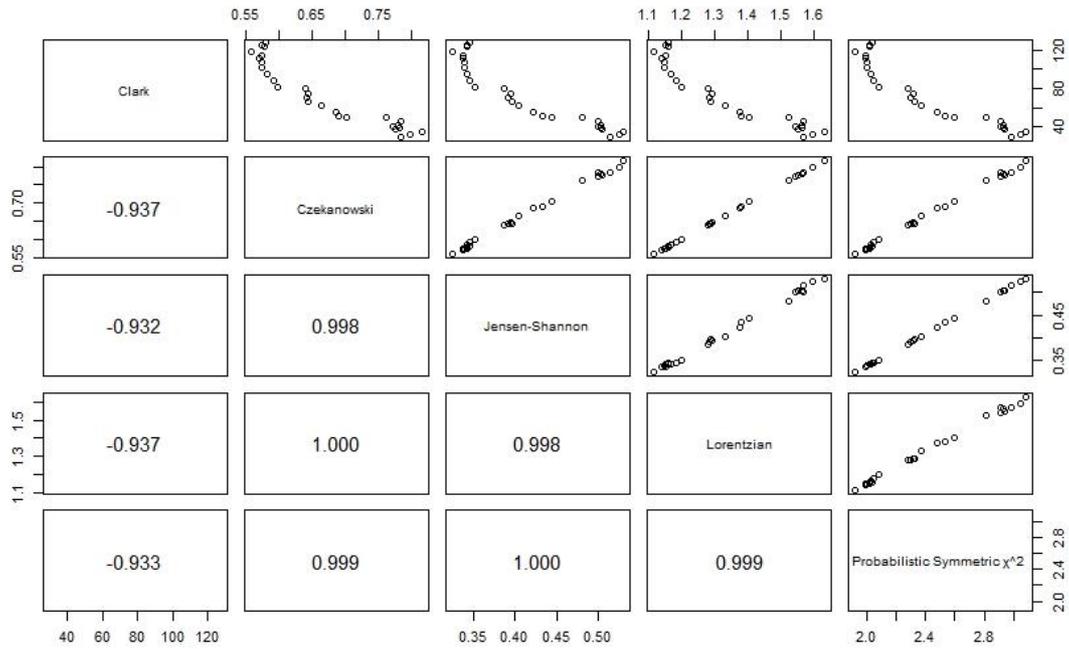

Figure 2. The scatterplot matrix among the Clark, Czekanowski, Jensen-Shannon, Lorentzian, and Probabilistic Symmetric $\chi^2$ measurements between the keyword distributions for each of these 28 successive pairs of years from 1991-2019 in the field of Environmental Science

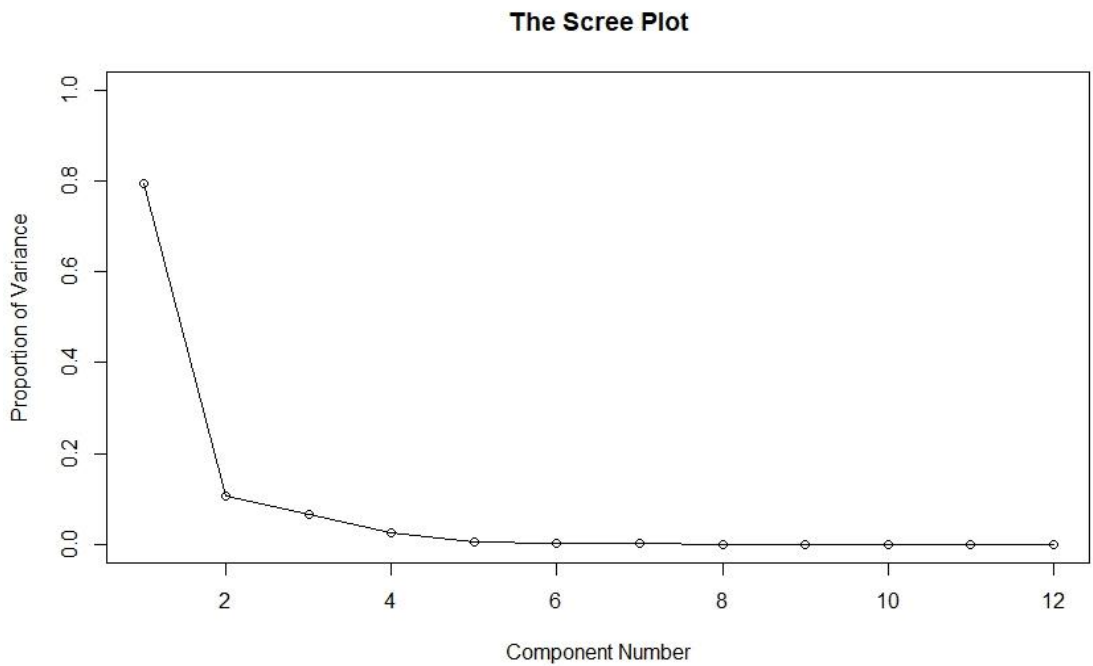

Figure 3. The scree plot of the principle component analysis for these 12 dissimilarity measures.

That is, the evolution between two successive years became smaller and smaller as time went by during this time period. The reason for this decreasing trend may be due to the fact that the research activities in two successive years have stronger dependence and connections and hence more similar keyword distributions in recent years than in the past. Second, the evolution measurements exhibit rises and falls that are not of a fixed frequency for Religion, Mechanical Informatics, and Information Science and Library Science (especially from 1991 to 2010), while there is a steady decrease for Environmental Science. These fluctuations may result from the transitions of research trends observed in these specific years. Third, the order by the amount of evolution among them changes over time. Religion have the largest value of evolution measurements among these four scientific fields most of the years from 1991 to 2009. Mechanical Informatics became the field with most evolution from 2009 to 2016 and then replaced by Information Science and Library Science afterwards.

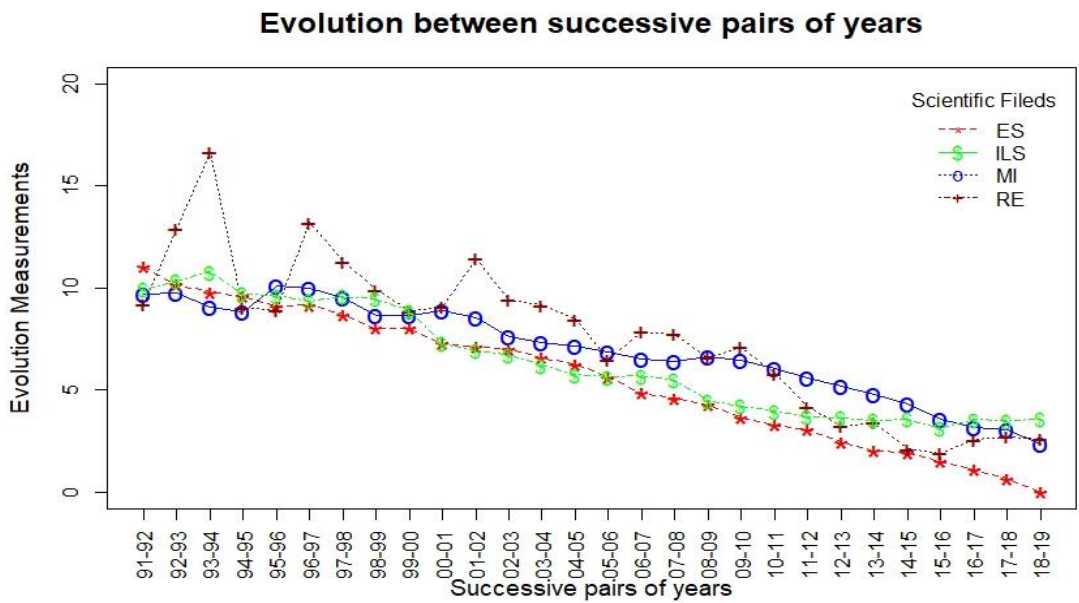

Figure 4. The evolution measurements for each successive pair of years from 1991 to 2019 in four chosen scientific fields: Environmental Science (ES), Information Science and Library science (ILS), Medical Informatics (MI), and Religion (RE).

Table 7 presents the average speed of evolution defined in Equation (16) during different time periods (1991-2000, 2001-2010, 2011-2019, and all years) in these four chosen scientific fields: Environmental Science (ES), Information Science and Library science (ILS), Medical Informatics (MI), and Religion (RE). During the period from 1991 to 2000, Religion ranked first in the evolution speed with an average of 11.121 per year, followed by Information Science and Library science (9.765 per year), Medical Informatics (9.366 per year) and Environmental Science (9.304 per year). From 2001 to 2010, Religion remains at the first place with an average of 8.347 per year, while Medical Informatics took the second place narrowing the gap between itself to Religion with an average of 7.243 per year. Information Science and Library Science dropped to the third place and Environmental Science remains at the bottom. From 2010 to 2019, Medical Informatics became the first in evolution speed with an average speed of 4.235 per year; Information Science and Library science became the second with an average of 3.575 per year; Religion dropped to the third place with an average speed of 3.173 per year; Environmental Science remains at the bottom If we consider all the years from 1991 to 2019, Religion ranks the first in evolution speed with an average of 7.575 per year, followed by Medical Informatics (6.958 per year), Information Science and Library science (6.375 per year), and Environmental Science (5.612 per year).

TABLE 7. The average speed of evolution defined in Equation (16) during different time periods (1991-2000, 2001-2010, 2011-2019, and all years) in these four chosen scientific fields: Environmental Science (ES), Information Science and Library science (ILS), Medical Informatics (MI), and Religion (RE).

|     | **1991-2000** | **2001-2010** | **2010-2019** | **All Years** |
| --- | --- | --- | --- | --- |
| **ES** | 9.304 | 5.735 | 1.783 | 5.612 |
| **ILS** | 9.756 | 5.852 | 3.575 | 6.375 |
| **MI** | 9.366 | 7.243 | 4.235 | 6.958 |
| **RE** | 11.121 | 8.347 | 3.173 | 7.575 |

**Conclusion**

In this paper, we developed a method to measure the evolution in a scientific field combing a collection of dissimilarity measures using principal component analysis, while most of the existing studies use only few (mostly just one single) dissimilarity measures. The problem with these works is that they failed to discuss the comparison and choices of such measures, ignoring the fact that there are many different dissimilarity measures. To the best knowledge of the authors, this is one of the first efforts in literature on finding the appropriate measures among such a collection of choices in the study of scientific evolution.

We chose four scientific fields: environmental studies, information science & library science, mechanical informatics, and religion to demonstrate the usage of the proposed methodology. We conducted a thorough analysis and comparison of different textual dissimilarity measures and performed a systematic analysis on how these scientific fields evolve and compare their evolution speeds over different time periods in the last three decades.

Note that, in this work, though the twelve dissimilarity measures are carefully selected and integrated in our methodology, they are in no way the best collection of dissimilarity measures possible. Instead, we believe that more efforts are needed to systematically study various properties of different dissimilarity measures in the study of scientific evolution. We merely seek to demonstrate the necessity and one approach to integrate different dissimilarity measures together in measure scientific evolution. Future work is needed to address the limitations and advantages of this approach.

**References**

Arnold, M. L. (1997). *Natural hybridization and evolution.* Oxford University Press on Demand.

Bateson, G. (2000). *Steps to an ecology of mind: Collected essays in anthropology, psychiatry, evolution, and epistemology.* University of Chicago Press.

Cai, Y., Xie, H., Lau, R. Y., Li, Q., Wong, T. L., & Wang, F. L. (2019). Temporal event searches based on event maps and relationships. *Applied Soft Computing, 85*, 105750.

Cha, S. H. (2007). Comprehensive survey on distance/similarity measures between probability density functions. *City, 1*(2), 1.

Chen, B., Tsutsui, S. Ding, Y., & Ma, F. (2017). Understanding the topic evolution in scientific domain: An exploratory study for the field of information retrieval. *Journal of Informetrics, 11*(4), 1175-1189.

Cleasby, I. R., Wakefiled, E. D., Morrissey, B. J., Bodey, T. W., Votier, S. C., Bearhoo, S., & Hamer, K. C. (2019). Using time-series similarity measures to compare animal movement trajectories in ecology. *Behavioral Ecology and Sociobiology, 73*(11), 151.

Darwin, C., & Bynum, W. F. (2009). *The origin of species by means of natural selection: or, the preservation of favored races in the struggle for life.* Harmondsworth: Pengiun.

Dias, L., Gerlach, M., Scharloth, J., & Altmann, E. G. (2018). Using text analysis ot quantify the similarity and evolution of scientific disciplines. *Royal Society open science, 5*(1), 171545.

Donald, M. (1991). *Origins of the modern mind: Three stages in the evolution of culture and cognition.* Harvard University Press.

Duda, R. O., Hart, P. E., & Stork, D. G. (2012). *Pattern classification.* John Wiley & Sons.

Finardi, U. (2014). On the time evolution of received citations, in different scientific fields: An empirical study. *Journal of Informetrics, 8*(1), 13-24.

Fordham, M. (2019). *Britain's Trade and Agriculture: Their Recent Evolution and Future Development.* Routledge.

Frank, M. R., Wang, D., Cebrian, M., & Rahwan, I. (2019). The evolution of citation graphs in artificial intelligence research. *Nature Machine Intelligence, 1*(2), 79-85.

Gaul, W., & Vincent, D. (2017). Evaluation of the evolution of relationships between topics over time. *Advances in Data Analysis and Classification, 11*(1), 159-178.

Giavazzi, F., Petkov, I., & Schiantarelli, F. (2019). Culture: Persistence and evolution. *Journal of Economic Growth, 24*(2), 117-154.

Grunspan, D. Z., Nesse, R. M., Barnes, M.E., & Brownell, S. E. (2018). Core principles of evolutionary medicine: a Delphi study. *Evolution, medicine, and public health*, 13-23.

Jain, G. Mahara, T., & Tripathi, K. N. (2020). A Survey of Similarity Measures for Collaborative Filtering-Based Recommender System. (Springer, Ed.) *Soft Computing: Theories and Applications*, 343-352.

Khan, M. Z., Mishra, A., & Khan, M. H. (2020). *Cyber Forencisc Evolution and Its Goals.* In Critical Concepts, Standards, and Techniques in Cyber Forensics (pp 16-30). IGI Global.

Langs, R. (2019). *The evolution of the emotion-processing mind.* Routledge.

McCulloch, J., & Wagner, C. (2020). On the choice of similarity measures for type-2 fuzzy sets. *Information Sciences, 510*, 135-154.

Mueller, U. G., Rehner, S. A., & Schultz, T. R. (1998). The evolution of agriculture in ants. *Science, 281*(5385), 2034-2038.

Nesse, R. M. (2008). The importance of evolution for medicine. *Evolutionary Medicine*, 416-432.

Ohno, S. (2013). *Evolution by gene duplication.* Springer Science & Business Media.

Panksepp, J., & Panksepp, J. B. (2000). The seven sins of evolutionary psychology. *Evolution and cognition, 6*(2), 108-131.

Poore, J., Flores, J. C., & Atkison, T. (2013). Evolution of digital forensics in virtualization by using virtual machine introspection. *In Proceedings of the 51st ACM Southeast Conference*, (pp. 1-6).

Shlens, J. (2014). A tutorial on pricipal component analysis. *arXiv preprint arXiv:1404.1100.*

Tang, S. (2020). *On Social Evolution: Phenomenon and Paradigm.* Routledge.

Tang, X., Yang, C., & Song, M. (2013). Understanding the evolution of multiple sceintific research domains using a content and network approach. *Journal of the American Society for Information Science and Technology, 64*(5), 1065-1075.

Vargas-Quesada, B., de Moya-Anegon, F., Chinchilla-Rodriguez, Z., &Gonzalez-Molina, A. (2010). Showing the essential science structure of a scientific domain and its evolution. *Informaiton Visualization, 9*(4), 288-300.

*Web of Science*. (n.d.). Retrieved from https://clarivate.com/webofsciencegroup/

Webb, A. R. (2003). *Statistical pattern recognition.* John Wiley &Sons.

Wei, G. (2018). Some similarity measures for picture fuzzy sets and their applications. *Iranian Journal of Fuzzy Systems, 15*(1), 77-89.

Xu, J., Bu, Y., Yang, S., Zhang, H., Yu, C., & Sun, L. (2018). Understanding the formation of interdisciplinary research from the perspective of keyword evolution: a case study on joint attention. *Scientometrics, 117*(2), 973-995.

Xu, S., Hao, L., An, X., Yang, G., & Wang, F. (2019). Emerging research topics detection with multiple machine learning models. *Journal of Informetrics, 13*(4), 100983.

Xu, Y., Zhang, W., Yang, S., & Shen, Y. (2019). Research Front Detection and Topic Evolution Based on Topological Structure and the PageRank Algorithm. *Symmetry, 11*(3), 310.

Zheng, H., & Zheng, L. (2020). An Investigation on Language Programs in US Higher Institutions—A Case Study on Chinese Language Programs. US-China Education Review, 10(6), 257-265.

Zhou, H., Yu, H., Hu, R. & Hu, J. (2017). A survey on trends of cross-media topic evolution map. *Knowledge-Based Systems, 124*, 164-175.